\documentclass[]{dcthesis}

\usepackage{amsmath,amssymb,amsthm}
\usepackage{graphicx}
\usepackage{fourier}
\usepackage{deluxetable}
\usepackage{booktabs}
\usepackage{amsmath}
\usepackage{float}
\usepackage{rotating}
\usepackage{lscape}
\usepackage{indentfirst}
\usepackage{natbib}
\usepackage{physics}
\usepackage{slashed}
\usepackage{tikz}

\usepackage[compat=1.1.0]{tikz-feynman}
\everymath{\displaystyle}

\def\apjl{Astrophys . J . Lett .}
\def\apjs{Astrophys . J . Supp . Ser .}

\def\aap {Astron . Astrophys .}

\theoremstyle{definition}

\theoremstyle{remark}

\title{Cold Dark Matter and Dark Energy Based on an Analogy with Superconductivity}
\author{Guanming Liang}
\date{\today}
\field{Physics}
\degree{Bachelor of Arts}
\committee{Robert R. Caldwell}{Miles P. Blencowe}{Rufus Boyack}{}

\begin{document}
\fontfamily{ptm}\selectfont

\frontmatter

\maketitle

\chapter*{Abstract}

We present a novel candidate for cold dark matter consisting of condensed Cooper pairs in a theory of interacting fermions with broken chiral symmetry. Establishing the thermal history from the early radiation era to the present, the fermions are shown to behave like standard radiation at high temperatures, but then experience a critical era decaying faster than radiation, akin to freeze-out that sets the relic abundance. Through a second-order phase transition, fermion - antifermion pairs condense and the system asymptotes towards zero temperature and pressure. By the present era, the non-relativistic, massive condensate decays slightly faster than in the standard scenario -- a unique prediction that may be tested by combined measurements of the cosmic microwave background and large scale structure. We also show that in the case of massive fermions, the phase transition is frustrated, and freeze-out instead leaves a residual, long-lived source of dark energy. 

\setcounter{page}{4}
\addcontentsline{toc}{section}{Abstract}

\chapter*{Preface}
\section*{Acknowledgements}
\addcontentsline{toc}{section}{Preface}

This thesis would not have been possible without the support of my family, friends, and teachers. I was extremely fortunate to have met my advisor Robert Caldwell as early as my first undergraduate quarter at Dartmouth, when I sat in on his course "PHYS 77: Introduction to General Relativity", sparking my curiosity in cosmology and gravity. In sophomore summer, we began working on the project that turned into this thesis, after taking "PHYS 92: Physics of the Early Universe". During junior year, we convened several times per week and my naive questions as a novice researcher were met with Robert's extraordinary patience. I especially appreciate his vast reservoir of knowledge in cosmology, his patient guidance as my advisor, and his critical insight that led us out of the misty fog. 

After Robert, the next person whom I appreciate the most is Miles Blencowe, with whom I worked on my first-ever research project in freshman year. He helped me gain a taste for theoretical research and inspired me to keep exploring new possibilities when old paths are obstructed. Miles has always been someone from whom I seek reliable advice in physics, life, and my future.

I would also like to acknowledge the people who critically influenced me. I want to thank Roberto Onofrio, with whom I took elementary quantum physics and "PHYS 72: Introductory Particle Physics". I am grateful to Devin Walker for teaching me the basics of quantum field theory in "PHYS 107: Relativistic QFT". I sat in on Rufus Boyack's "PHYS 113: Solid State Physics", where I learned about condensed matter field theory and superconductivity. Together with Rufus, we figured out parts of the effective field theory regularization scheme in this thesis. I also wish to express special gratitude to Yuhang Zhu and Yi Wang at the Hong Kong University of Science and Technology. Their paper \textit{BCS in the Sky} motivated my interest in the subject matter. 

Last but certainly not least, I wish to acknowledge my dear parents, who encourage me to pursue my academic dream with unanimous support. Thank you to Ashley Wang '25, for all her love and support. Thank you to Tunmay Gerg '25 and Bradyn Quintard '25, with whom I worked on problem sets and prepared for exams in various physics and math courses. Thank you to Shadi Ali Ahmad '22, who encouraged me to pursue theoretical research. Thank you to Nat Alden '23 and Matthew Goodbred '23 for working with me on P92 problem sets and giving me great advice to remember. All of you have truly shaped my character and knowledge.

\tableofcontents


\listoffigures

\mainmatter

\chapter{Introduction}

The material presented in this thesis is original and initially appeared in Phys. Rev. Lett. \textbf{134}, 191004 (2025) \citep{Liang:2024xww}. Due to the Letter's word limit, this thesis comprehensively expands on the mathematical derivations and devotes more explanations to the subtle physics. We also show additional figures for illustration purposes.  

The $\Lambda$CDM cosmological model is well supported by a wide range of observational and experimental data \citep{Planck:2018vyg, BOSS:2013rlg,DES:2021wwk,Brout:2022vxf,DESI:2024mwx}. Yet, the nature of its two main ingredients, approximately $25\%$ cold dark matter (CDM) and $70\%$ cosmological constant-like dark energy ($\Lambda$), remains elusive to our understanding. 

The leading paradigm for dark matter posits a cold, non-relativistic particle species that falls out of equilibrium with the primordial thermal bath as the Universe expands. The freeze out mechanism and observed relic abundance of dark matter imply an electroweak energy scale for the interaction with the Standard Model (SM). However, laboratory searches at these scales for weakly-interacting massive particles (WIMPs) have produced null results to date \citep{Roszkowski:2017nbc}. Non-thermal relics including axions \citep{Chadha-Day:2021szb} and primordial black holes \citep{Carr:2016drx} are actively sought through direct experimental detection \citep{ADMX:2018gho,Adams:2022pbo} and astrophysical observations \citep{DeLuca:2020agl,Bartolo:2018rku} but remain at large \citep{Baer:2014eja,Carr:2020xqk}. The physics of dark matter is an open question.

The conventional explanation for the accelerated expansion of the Universe is a cosmological constant, $\Lambda$. However, naive estimates of the quantum vacuum energy are famously many orders of magnitude different from observation \citep{Weinberg:1988cp}, and $\Lambda$ is widely regarded as a placeholder until a deeper understanding of dark energy can be achieved \citep{Carroll:2000fy}. Though recent results from the Dark Energy Spectroscopic Instrument (DESI) are challenging the cosmological constant \citep{DESI:2025zgx}, future cosmological surveys are critical for further understanding of dark energy, the fluid nature of which remains unknown. 

The organization of this thesis is as follows. In Chap.\ref{njlmodel}, we investigate the Nambu-Jona-Lasinio (NJL) model of interacting fermions \citep{Nambu:1961tp,Nambu:1961fr}. This theory was originally proposed as a relativistic model for dynamical mass generation, in analogy to non-relativistic Bardeen-Cooper-Schrieffer (BCS) superconductivity. More recently, the NJL model has been used in cosmological models of inflation \citep{Tong:2023krn}, dark matter \citep{Alexander:2016glq,Alexander:2024qml,Garani:2022quc,Alexander:2018fjp,Alexander:2020wpm}, dark energy \citep{Alexander:2006we,Alexander:2009uu,Inagaki:2003qq}, and more \citep{Tukhashvili:2023itb,Alexander:2008vt}. 

In Chap.\ref{finiteT}, we explain the critical role of the NJL thermal history. In most prior investigations under cosmological contexts, NJL fermions were either assumed to be at zero temperature or in equilibrium with the thermal bath of SM particles. That is, the distinct thermal behavior of NJL fermions was overlooked. 

In Chap.\ref{dmde}, we propose novel candidates for CDM and dark energy as two different cases of a system of interacting fermions with broken chiral symmetry, similar to the condensation of Cooper pairs in BCS superconductivity \citep{Bardeen:1957mv}.

In brief, we present a scenario in which the relativistic NJL fermions decouple from the SM at early times, evolving like radiation with separate temperature from the thermal bath of SM particles. When the temperature drops past the axial chemical potential, the interacting fermion system begins to stiffen with an effective equation of state $w \equiv p/\rho = 1$, defined as the ratio of the homogeneous pressure to energy density. The duration of this stage sets the relic dark matter abundance, similar to the role of freeze out for thermal dark matter. Below a critical temperature, the system undergoes a phase transition to the true vacuum, and the fermions coalesce into a cold, massive condensate as CDM. The standard Big Bang cosmology with dark matter proceeds as expected, from approximately GeV scales on down in the example we present.

In a second scenario, massive fermions are shown to spoil the phase transition, and the condensate is stuck in a long-lived, metastable vacuum as dark energy. The thermal history dictates that the equation of state  proceeds in stages as $w= 1/3\,\rightarrow\, 1\, \rightarrow\, -1$. In this view, the fine tuning and coincidence problems are resolved, similar to models of quintessence \citep{Zlatev:1998tr}; the dark energy is set by the frozen dynamics of the condensate.


\chapter{Particle Model}\label{njlmodel}

\section{NJL Theory}
The proposed NJL theory of CDM (and dark energy) is inspired by analogy with BCS superconductivity, consisting of fermion pairs $\bar\psi\psi$ each with Dirac mass $m$ and a quartic, scalar, attractive self-interaction with coupling $M>0$: 
\vspace{-1em}
\begin{align}\label{eqn:dirac}
\mathcal{L}&=\bar{\psi}\left(i\gamma^{\mu}\partial_{\mu}-m\right)\psi-\kappa\bar{\psi}\gamma^0\gamma^5\psi+\frac{\left(\bar{\psi}\psi\right)^2}{M^2}.
\end{align}
Here $\kappa$ is the axial chemical potential associated with an asymmetry between left- and right-handed fermions: $n_5=n_L-n_R = \langle\bar\psi\gamma^{\mu}\gamma^5\psi\rangle \neq 0$. Its origin can be prescribed to a baryogenesis scenario that takes place above the valid energy scale of this model. Possible scenarios include an electroweak phase transition with handedness, primordial chiral magnetic fields, and chiral gravitational waves. Since chemical potential $\mu$ is zero, the $U(1)$ vector charge $\langle \bar\psi\gamma^0\psi \rangle=n-\bar n=0$ is neutral by construction. The Cooper pairs in this theory are fermion-antifermion pairs of like helicity, equal and opposite momenta, which are drawn together by the self-interaction. Since the Hubble expansion rate is negligible compared to the energy scales and parameters of the theory, we are justified to work in flat space, using temperature as our clock.

Without axial asymmetry, the theory is effectively the low-momentum limit ($p^2\ll M^2$) of the Yukawa interaction, where $M$ is interpreted as the mediating boson mass. We note similarities and differences between our NJL theory (scalar-exchanging) and the four-Fermi interaction (with V-A structure). The latter is reproduced in the low-momentum limit ($p^2\ll M_W^2$, the W-boson mass) of the weak interaction. Both are contact interactions with a scattering rate of $\Gamma_{\rm scatter}=\langle n\sigma v\rangle\sim G_F^2T^5$, where $G_F=\frac{1}{M_W^2}$ is the weak coupling constant. Below, we illustrate a boson mediated scattering process (left) and low-momentum contact scattering (right). \\

\vspace{-1.2cm}
\begin{center}
\begin{minipage}{0.45\textwidth}
    \centering
    \includegraphics[width=\textwidth]{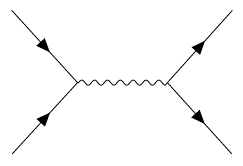}
\end{minipage}
\hspace{0\textwidth} 
\begin{minipage}{0.45\textwidth}
    \centering
    \includegraphics[width=\textwidth]{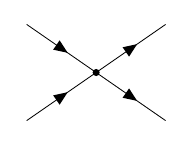}
\end{minipage}
\end{center}

The nonzero $\kappa$ construction is justified as follows. The chirality flipping rate $\Gamma_f$ defines the rate at which chiral asymmetry is erased. There are two potential sources of chirality flipping in the Lagrangian, $m\bar\psi\psi$ and $\frac{(\bar\psi\psi)^2}{M^2}$, because they contain chiral mixing terms $\bar\psi\psi=\bar\psi_L\psi_R+\psi_L\bar\psi_R$. Chirality flipping from $m\bar\psi\psi$ is negligible due to the small mass that concerns us. By analogy with weak interaction, the scattering cross section is $\sigma = \frac{T^2}{M^4}$ and the scattering rate $\Gamma = \langle n\sigma v\rangle$. In the context of cosmology, since $n\propto a^{-3}\propto T^3$ and $v=c=1$, $\Gamma=\frac{T^5}{M^4}$ coincides with the helicity flipping rate. Intuitively, since chirality is equivalent to the time-invariant (conserved) fermion helicity in the $m\rightarrow 0$ limit, a mass factor relates a definite helicity state with a definite chirality state. (In Appendix C, we prove helicity is conserved in our scalar-exchanging interaction.) For this reason, the quartic interaction has $\Gamma_f=\left(\frac{m}{3T}\right)^2\left(\frac{T^5}{M^4}\right)$ \citep{Pavlovic:2016mxq, Boyarsky:2020cyk}. For subsequent considerations of CDM and dark energy, the mass is either zero or small and suppresses chirality flipping. Therefore, $\Gamma_f$ is either zero or falls below the Hubble expansion rate $H$ sufficiently early, and the axial asymmetry persists. 

To better understand this theory, we write down the fermionic path integral $\mathcal{Z}$:
\begin{equation}
    \mathcal{Z}=\int D\bar{\psi}D\psi e^{i\int d^4x\mathcal{L}}
\end{equation}
where $\mathcal{L}$ is from Eq.(\ref{eqn:dirac}). Since it is beneficial to work with scalar fields in cosmology, we now introduce an auxiliary scalar field, the order parameter $\Delta$, which allows us to integrate out the fermions and rewrite Eq.(\ref{eqn:dirac}) as a scalar theory. Physically, $\Delta$ is interpreted as the energy gap. With the Hubbard-Stratonovich transformation \citep{hubbard1959calculation}, we rewrite the identity operator:
\begin{equation}
    1=\int D\Delta\exp \left[ -\int d^4x \left(\frac{1}{2}{M\Delta}-\frac{1}{M}{\bar{\psi}\psi}\right)^2 \right]
\end{equation}
Thus, we arrive at
\begin{equation}
    \int D\Delta\exp{i\int d^4x\left(\frac{\bar{\psi}\psi}{M}\right)^2}=\int D\Delta\exp{i\int d^4x \left(\bar{\psi}\psi\Delta-\frac{M^2\Delta^2}{4}\right)}.
\end{equation}
The fermion theory can be rewritten with a Grassman spinor property
\begin{equation}
    \int D\bar\psi D\psi\exp(i\int d^4x\bar\psi\mathcal{G}^{-1}\psi)=\det(\mathcal{G}^{-1}).
\end{equation}
Here, the inverse Green's function is the Fourier space equation of motion:
\begin{equation}
    \mathcal{G}^{-1}=-\gamma^{\mu}k_{\mu}+\Delta-m-\kappa\gamma^0\gamma^5
\end{equation}
Working in Fourier space $\partial_{\mu}\rightarrow ik_{\mu}$:
\begin{equation}
    \mathcal{L}=-\frac{1}{4}M^2\Delta^2+\int\frac{d^4k}{(2\pi)^4}\ln\det\mathcal{G}^{-1}, 
\end{equation}
Redefining $\Delta-m\rightarrow\Delta$, the tree-level contribution with one-loop corrections are
\begin{equation}\label{one-loop}
    \mathcal{L}= -\frac{1}{4}M^2(\Delta+m)^2+\int\frac{d^4k}{(2\pi)^4}\ln U_+ U_-, 
\end{equation}
where $U_{\pm} = \Delta^2-\omega^2+(k\pm\kappa)^2$. The corresponding path integral is now:
\begin{equation}
    \mathcal{Z}=\int D\Delta \exp{i \int d^4x \left[-\frac{1}{4} M^2 \Delta^2 -i\ln Z_f(\Delta)\right]},
\end{equation}
where $\ln Z_f(\Delta)=\int d^4x\int\frac{d^4k}{(2\pi)^4}(\ln U_+ + \ln U_-)$. 
\section{Zeta Regularization}
The integral in Eq.(\ref{one-loop}) is now explicitly evaluated, first over energy and then momentum. We treat the energy integration by Wick rotating $\omega\rightarrow i\omega_E$:
\begin{equation}
    \epsilon = i \int \frac{d\omega_E}{2 \pi} \left[ \ln\left(\Delta^2 + \omega_E^2 + (k-\kappa)^2\right) + (\kappa \to -\kappa) \right].
\end{equation}
This integral is formally divergent, but we can Zeta regularize it with a mathematical device:
\begin{equation}
    \epsilon = -i \left(\frac{d}{ds} \int \frac{d\omega_E}{2 \pi} \left[\Delta^2 + \omega_E^2 + (k-\kappa)^2\right]^{-s} + (\kappa \to -\kappa) \right)_{s=0}.
\end{equation}
Recall the integral definition of the $\Gamma(s)$ function for Re$(s)>0$ \citep{sebah2002introduction}:
\begin{equation}
    \Gamma(s)=\int_0^\infty dt  \, t^{s-1}{\rm e}^{-t}.
\end{equation}
We observe that under transformation of variables $t=ux$: 
\begin{equation}
    \int_0^\infty dt  \, t^{s-1}{\rm e}^{-tx} = x^{-s}\int_0^\infty \, du \, u^{s-1} \, {\rm e}^{-u}\, \rightarrow x^{-s} = \frac{1}{\Gamma(s)} \int_0^\infty dt \, t^{s-1} \, {\rm e}^{-t x}.
\end{equation}
Therefore, the energy integral becomes:
\begin{equation}
    \epsilon = -i\left( \frac{d}{ds} \, \frac{1}{\Gamma(s)}\, \int \frac{d\omega_E}{2 \pi} \int_0^\infty dt\, t^{s-1}\left[ {\rm e}^{-t(\Delta^2 + \omega_E^2 + (k-\kappa)^2)} + (\kappa \to -\kappa)\right]\right)_{s=0}.
\end{equation}
The Gaussian integration result 
\begin{equation}
    \int_{-\infty}^{\infty} \frac{d\omega_E}{2 \pi} {\rm e}^{-t \omega_E^2} = \frac{1}{2 \pi} \sqrt{\frac{\pi}{t}}
\end{equation}
enables us to simplify this integral:
\begin{eqnarray}
    \epsilon &=& -i\left( \frac{d}{ds} \, \frac{1}{2 \sqrt{\pi}\Gamma(s)}\, \int_0^\infty dt\, t^{s-3/2}\left[ {\rm e}^{-t(\Delta^2 + (k-\kappa)^2)} + (\kappa \to -\kappa)\right]\right)_{s=0} \cr
    &=& -\frac{i}{2 \sqrt{\pi}}\left( \frac{d}{ds} \frac{\Gamma(s-\tfrac{1}{2})}{\Gamma(s)} \left[ \left(\Delta^2 + (k-\kappa)^2\right)^{\tfrac{1}{2}-s} + (\kappa \to -\kappa)\right] \right)_{s=0},
\end{eqnarray}
where in the second equality, we used the identity 
\begin{equation}
    \int_0^\infty dt \, t^{s-3/2} {\rm e}^{-t x} = \Gamma(s - \tfrac{1}{2}) \, x^{-(s-1/2)}.
\end{equation}
Approximating $\epsilon$ at leading order for small values of $s>0$,
\begin{equation}
    \frac{d}{ds} \left(\frac{\Gamma(s-\tfrac{1}{2})}{\Gamma(s)} x^{\tfrac{1}{2}-s}\right) = \frac{d}{ds}\left(-2\sqrt{\pi} s + {\cal O}(s^2)\right),
\end{equation}
we find
\begin{eqnarray}
    \epsilon &=& i \left( \sqrt{\Delta^2 + (k-\kappa)^2} + \sqrt{\Delta^2 + (k+\kappa)^2}\right) \\
    A &=& \frac{i}{2 \pi^2} \int_0^\infty k^2 \, dk \, \left( \sqrt{\Delta^2 + (k-\kappa)^2} + \sqrt{\Delta^2 + (k+\kappa)^2}\right) \\
    L_{\rm eff} &=& -\frac{1}{4} M^2 \Delta^2 - iA. \label{eqn:Leff}
\end{eqnarray}
\section{Dimensional Regularization}
$A$ is another formally divergent loop integral. Let us define
\begin{equation}
    I_{\pm} = \int_0^\infty dk \, k^2 \, \sqrt{\Delta^2 + (k \pm \kappa)^2}
\end{equation}
and change variables $u_{\pm}=k\pm\kappa$. We get $I_++I_-=2\left(J_2+\kappa^2 J_0\right)$, where 
\begin{align}
    J_n=\int_0^\infty duu^n\sqrt{\Delta^2+u^2}.
\end{align}
Per dimensional regularization, we let the number of spacetime dimensions $D$ approach $4$ from below. Using Eq.(11.A.2) of Weinberg's Quantum Field Theory Volume 1 \citep{weinberg1995quantum}:
\begin{eqnarray}\label{eqn:dimreg}
    J_2 &=& \mu^{4-D} \int dk k^{D-2} \left.\sqrt{\Delta^2 + k^2} \right|_{D=4-\epsilon} \cr
    &=&\mu^{4-D} \Delta^D \left.\frac{\Gamma(\tfrac{D-1}{2})\Gamma(- \tfrac{D}{2})}{2 \Gamma(-\tfrac{1}{2})}\right|_{D=4-\epsilon} \\
    &=& -\Delta^4 \left( \frac{\mu}{\Delta} \right)^\epsilon \frac{\Gamma(\tfrac{3-\epsilon}{2})\Gamma(-2 + \tfrac{\epsilon}{2})}{4 \sqrt{\pi}} = -\frac{\Delta^4}{8} \frac{1}{\epsilon} + \frac{\Delta^4}{32}\left(1 + 4 \ln\frac{\Delta}{2 \mu}\right) + {\cal O}(\epsilon),
\end{eqnarray}
By a similar procedure:
\begin{equation}
    J_0=\frac{\Delta^2}{2}\frac{1}{\epsilon} + \frac{1}{4} \Delta^2 \left(1 - 2 \ln \frac{\Delta}{2 \mu}\right) + {\cal O}(\epsilon).
\end{equation}
Since this is a low-momentum effective field theory, we can absorb $\epsilon$ and $\mu$ into an ultraviolet cutoff $\Lambda_{UV}$, whereby $\frac{1}{\epsilon} + \ln 2 \mu = \ln \Lambda$. To leading order, we find:
\vspace{-.3em}
\begin{equation}
    V_{\rm eff}\left(\Delta\right) = \frac{1}{4}M^2(\Delta+m)^2 + \frac{\Delta^4}{32 \pi^2}\left(1 + 4 \frac{\ln{\Delta}}{\Lambda_{UV}}\right) - \frac{\kappa^2\Delta^2}{4 \pi^2}\left(1 - 2 \frac{\ln\Delta}{\Lambda_{UV}}\right).
    \label{eqn:vloop}
\end{equation}
where $V_{\rm eff}(\Delta)=-L_{\rm eff}(\Delta)+\mathcal{O}(\partial_{\mu}\Delta)^2$ is not quite the effective potential. 

\section{Effective Potential}
In cosmology where negative pressure drives the expansion, we lift the effective potential by $V_0$ so that the non-thermal $V_{\rm eff}$ is zero at the minimum: 
\vspace{-.3em}
\begin{equation}\label{eqn:vloop}
    V_{\rm eff}\left(\Delta\right) = \frac{1}{4}M^2(\Delta+m)^2 + \frac{\Delta^4}{32 \pi^2}\left(1 + 4 \frac{\ln{\Delta}}{\Lambda_{UV}}\right) - \frac{\kappa^2\Delta^2}{4 \pi^2}\left(1 - 2 \frac{\ln\Delta}{\Lambda_{UV}}\right) + V_0.
\end{equation}
Eq.(\ref{eqn:vloop}) is one of the main results of this thesis. Self-consistency of this effective theory requires $M \gtrsim \Lambda_{UV} \gtrsim \kappa$, since $M$ is the tree-level mass scale. A striking feature of this model is the emergence of an exponentially-suppressed energy scale. For sufficiently small mass $m\ll\Delta_0$, there is an exponentially suppressed local minimum, the gap, at approximately 
\begin{equation} 
    \Delta_0 = \Lambda_{UV} \exp(-\frac{\pi^2M^2}{2\kappa^2}).
\end{equation} 
Since $M>\kappa$, $\Delta_0$ is small with respect to $\Lambda_{UV}$. 

\section{Axial Charge}
The chiral asymmetry number density is defined as $n_A=\langle\bar\psi\gamma^0\gamma^5\psi\rangle$. It is straightforward to calculate this in the path integral approach:
\begin{align}
    n_A&=\int D\bar\psi D\psi e^{\beta\int d^3x\mathcal{L}}\bar\psi\gamma^0\gamma^5\psi=\int D\Delta e^{\beta\int d^3x\mathcal{L}}\frac{\delta\ln\det\mathcal{G}^{-1}}{\delta\kappa}\\
    &=-\int D\Delta e^{\beta\int d^3x\mathcal{L}}\frac{\partial}{\partial\kappa}\left(V_{\rm eff}(\Delta)-\frac{M^2(\Delta+m)^2}{4}\right)=-\Biggl\langle\frac{\partial V_{\rm eff}}{\partial\kappa}\biggr\rangle
\end{align}

\section{Vector Charge}
The theory's lack of chemical potential $\mu$ ensures the U(1) vector charge density is zero. This is now verified by averaging the zeroth component of the vector current $\rho_V = \langle \bar\psi \gamma^0 \psi \rangle$. Here, the mathematical trick is to add a fictitious current $b\bar\psi \gamma^0 \psi$ to the Lagrangian:
\begin{equation}
    \mathcal{L}=\bar{\psi}\left(i\gamma^{\mu}\partial_{\mu}-m\right)\psi-\kappa\bar{\psi}\gamma^0\gamma^5\psi+\frac{\left(\bar{\psi}\psi\right)^2}{M^2}+b\bar\psi \gamma^0 \psi.
\end{equation}
The equation of motion is now:
\begin{equation}
    {\cal G}^{-1} = i \gamma^\mu \partial_\mu - \kappa \gamma^0 \gamma^5 + \Delta + b \gamma^0
\end{equation}
It thus follows that the determinant of ${\cal G}^{-1}$ is
\begin{equation}
    \ln{\rm det}{\cal G}^{-1} = \ln\left(\Delta^2 -(b+\omega)^2 + (k - \kappa)^2\right) + (\kappa \to -\kappa).
\end{equation}
Now, define $\omega' = \omega + b$ and since we integrate over $\frac{d^4k}{(2\pi)^4}$ with $\omega$ running from $-\infty$ to $\infty$, the integration result is not dependent on $b$. For this reason, $\rho_V=\left.\frac{\partial V_{\rm eff}}{\partial b}\right|_{b=0}=0$.

\section{The Pseudoscalar Interaction}
Instead of the scalar four-fermion interaction, we may consider a pseudoscalar coupling $(\bar\psi i\gamma^5 \psi)^2$. In this case, the inverse Green's function $\mathcal{G}^{-1}$ becomes
\begin{equation}
    \ln{\rm det}{\mathcal{G}^{-1}} = \ln\left(\Delta^2 + m^2 - \omega^2 + (k - \kappa)^2\right) + (\kappa \to -\kappa).
\end{equation}
Defining $\Delta'^2 = \Delta^2 + m^2$, the effective potential is
\begin{eqnarray}
    V_{\rm eff} = \frac{1}{4} M^2 (\Delta'^2 - m^2) + \frac{\Delta'^4}{32 \pi^2}\left(1 + 4 \ln\frac{\Delta'}{\Lambda}\right) - \frac{\kappa^2\Delta'^2}{4 \pi^2}\left(1 - 2 \ln \frac{\Delta'}{\Lambda}\right).
\end{eqnarray}
This is similar to but bears differences from Eq.(\ref{eqn:vloop}). For $m=0$, however, the pseudoscalar case is identical to the scalar case. 

\section{Pseudoscalar + Scalar Interaction}
It is natural to consider a pseudoscalar interaction in addition to the scalar interaction:
\begin{equation}
    L=\bar\psi(i\gamma^{\mu}\partial_{\mu}-m)\psi-\kappa\bar\psi\gamma^0\gamma^5\psi+\frac{(\bar\psi\psi)^2}{M^2}+\frac{(\bar\psi i\gamma^5\psi)^2}{M_5^2}.
\end{equation}
In parallel with the single-field scenario, we define:
\begin{align*}
    1&=\int D\Sigma\exp \left[ -\int d^4x \left(\frac{1}{2}{M\Sigma}-\frac{1}{M}{\bar{\psi}\psi}\right)^2 \right] \\ 
    1&=\int D\Pi\exp \left[ -\int d^4x \left(\frac{1}{2}{M_5\Pi}-\frac{1}{M_5}{\bar{\psi}i\gamma^5\psi}\right)^2 \right].
\end{align*}
where $\frac{1}{2}M^2\langle\Sigma\rangle=\langle\bar\psi\psi\rangle$ and $\frac{1}{2}M_5^2\langle\Pi\rangle=\langle\bar\psi i\gamma^5\psi\rangle$. Applying the Hubbard-Stratonovich technique separately to each quartic interaction,
\begin{equation}
    \ln\det\mathcal{G}^{-1}=\ln\left[\Sigma^2+\Pi^2-2\Sigma m+m^2-\omega^2+(k-\kappa)^2\right]\left[\Sigma^2+\Pi^2-2\Sigma m+m^2-\omega^2+(k+\kappa)^2\right].
\end{equation}
We now change variables to $\Delta=\sqrt{\Sigma^2+\Pi^2}$ and $\Theta=\arctan(\Pi/\Sigma)$. Since $\Delta^2-2m\Delta\cos\Theta+m^2\geq m^2\sin\Theta^2\geq 0$, we can define $\Delta_5^2=\Delta^2-2m\Delta\cos\Theta+m^2\geq 0$. Thus, the effective potential is:
\begin{eqnarray}
    V_{\rm eff}&=&\frac{M^2\cos^2\Theta+M_5^2\sin^2\Theta}{4}\left(m^2\cos2\Theta+\Delta^2_5+2m\cos\Theta\sqrt{\Delta^2_5-m^2\sin^2\Theta}\right) \cr
     &+&\frac{\Delta^4_5}{32\pi^2}\left(1+4\log\frac{\Delta_5}{\Lambda}\right)-\frac{\kappa^2}{4\pi^2}\left(1-2\log\frac{\Delta_5}{\Lambda}\right)\Delta^2_5
\end{eqnarray}
For $m=0$, $V_{\rm eff}$ changes at the tree-level but not at one-loop. If $M=M_5$, we lose the phase dependence on $\Theta$ and $V_{\rm eff}$ reduces to the original effective potential for the scalar interaction. We further note the existence of a Nambu-Goldstone boson in this theory, because the pseudoscalar interaction spontaneously breaks the chiral U(1) symmetry. Although this generalized case is interesting from the perspective of particle physics, we will focus on the single-field scenario in the subsequent discussion.

\chapter{Finite Temperature Field Theory}\label{finiteT}
\section{Formalism}
In the previous section, we have implicitly assumed the effective field theory is at zero temperature. To understand its unique thermal behavior, we calculate its corresponding finite temperature field theory. This is achieved through the Matsubara formalism, where we impose periodic (antiperiodic) boundary conditions on bosonic (fermionic) quantum fields over imaginary, Wick-rotated time. Their periodicity is equal to the inverse temperature $\beta=\frac{1}{T}$, which diverges to infinity at the $T=0$ limit. Therefore, in the Fourier transform to momentum space, we replace the $\omega$ integral of the zero temperature theory with a Matsubara summation over discrete fermionic modes, $\omega_n=\frac{(2n+1) \pi}{\beta}$,  $\frac{d\omega}{2\pi} \rightarrow \frac{1}{\beta}\sum_{i\omega_n}$ and $\mathcal{G}^{-1}\rightarrow \beta\mathcal{G}^{-1}$. The resulting one loop correction is thus
\vspace{-.3em}
\begin{equation}
    \mathcal{L}_1(\Delta, T) = \int\frac{d^3k}{(2\pi)^3}\frac{1}{\beta} \sum_{n=-\infty}^\infty  \ln\beta^4 U_{n+}U_{n-} \nonumber
\end{equation}
where $U_{n\pm} = \Delta^2 + \omega_n^2+(k\pm\kappa)^2$.
By the complex identity \citep{kapusta2007finite}
\begin{equation}
    \lim_{\delta\to 0}\sum_{n=-\infty}^{\infty} {\rm e}^{i\omega_n \delta} \ln(\beta s - i \omega_n \beta) = \ln(1 + {\rm e}^{-\beta s}),
\end{equation}
we sum over the fermionic frequencies:
\begin{equation}
    \mathcal{L}_1(\Delta, T)=\int\frac{d^3k}{(2\pi)^3}\frac{1}{\beta}\left[\ln(1+e^{\beta Q_-}) + \ln(1+e^{-\beta Q_-})+(Q_-\rightarrow Q_+)\right]\nonumber
\end{equation}
where $Q_{\pm} = \sqrt{\Delta^2 + (k \pm \kappa)^2}$. Notice that:
\begin{equation}
    \frac{1}{\beta}\ln(1 + {\rm e}^{\beta Q_{\pm}}) = \frac{1}{\beta}\ln({\rm e}^{\beta Q_{\pm}}(1 + {\rm e}^{-\beta Q_{\pm}})) = Q_{\pm} + \frac{1}{\beta}\ln(1 + {\rm e}^{-\beta Q_{\pm}}).
\end{equation}
The theory can be separated into a zero temperature contribution as well as finite temperature corrections:
\begin{equation}
     V_{\rm eff}(\Delta,\,T) = V_{\rm eff}(T=0) -\frac{2}{\pi^2\beta^4} [ I_2(\beta\Delta) + \kappa^2 \beta^2 I_0(\beta\Delta) ] \label{eqn:vt}
\end{equation}
where 
\begin{equation}
    I_n(x)=\int_{0}^{\infty}ds\, s^n\ln \left[ 1+e^{-\sqrt{s^2+x^2}} \right].
\end{equation}
Eq.(\ref{eqn:vt}) is important because it contains the unique thermal history of the system.

In Eq.(\ref{eqn:vt}), $I_0(0) = \pi^2/12$ and $I_2(0) = 7 \pi^4/360$. We can obtain high and low temperature approximations for the thermal contributions. At $T \gg \kappa,\, m,\, \Delta$, when the minimum of $V_{\rm eff}$ is at $\Delta=0$:
\begin{equation}\label{eqn:HighTVeff}
    V_{\rm eff} = -\frac{7}{8}\frac{2 \pi^2}{45} T^4.
\end{equation}
where $T^4$ is characteristic of cosmological radiation. At low temperatures, we approximate $\ln(1+e^{-\beta\Delta}) \simeq e^{-\beta\Delta}$ and $I_n(x)$ can be analytically integrated and written in terms of the modified-Bessel and Gamma functions. $I_0(x)=x K_1(x)$ and $I_2(x)=x^2 K_2(x)$, both of which decay to zero as $x\rightarrow\infty$. The leading terms in the finite-temperature potential are given by
\begin{equation}
    I_n(x \gg 1) \approx \sqrt{\frac{\pi}{2}} x^{\frac{n+1}{2}} {\rm e}^{-x},
\end{equation}
meaning finite temperature corrections to the potential are exponentially suppressed for $\Delta\gg T$. In the case $\kappa \gg \Delta \gg T$, the finite-temperature contribution is approximately
\begin{equation}
    -\frac{2}{\pi^2}\sqrt{\frac{\pi T}{2 \Delta}}{\rm e}^{-\Delta/T} \Delta \kappa^2 T,
\end{equation} 
where the leading term depends on the asymmetry through $\kappa$ and contributes negligibly to the potential.

\section{Second Order Phase Transition}
\subsection{Gap Equation}
\begin{figure}
    \centering
    \includegraphics[width=0.7\linewidth]{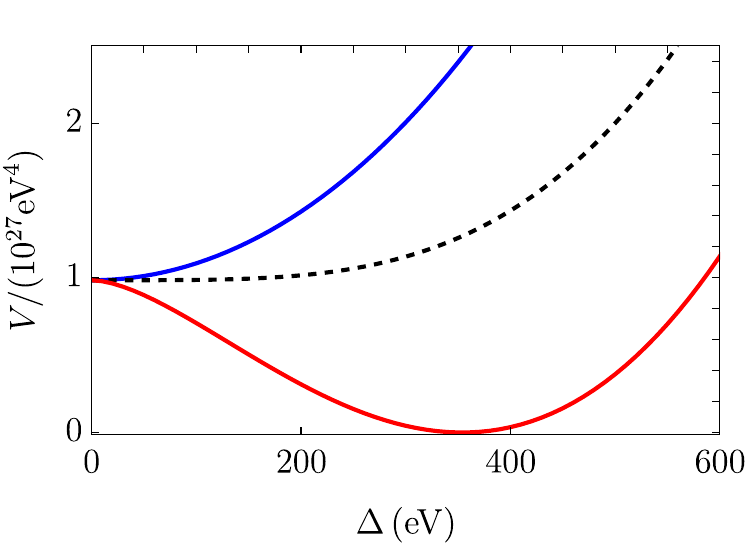}
    \caption{Effective potential $V(\Delta,T)$ versus $\Delta$ for the massless theory at $T=2 T_c$ (blue), $T=T_c$ (black, dashed), and $T=0$ (red). The parameters are $(\kappa,\,\Lambda_{UV},\,M)=(0.56, 1.1, 1.2)$~TeV, for the cold dark matter scenario. For illustrative purposes, the $T\neq 0$ curves have been artificially shifted vertically to match the $T=0$ curve at $\Delta=0$.}
    \label{fig:phase-transition}
\end{figure}

The thermal history is now clear from a numerical standpoint. 
Above a critical temperature, $T > T_c$, the gap field $\Delta$
sits at zero. For $T < T_c$ in the massless case the
field undergoes a second-order phase transition, where
the local minimum smoothly transfers from $\Delta = 0$ to $\Delta_0$. Fig.(\ref{fig:phase-transition}) illustrates this change in $V_{\rm eff}$ as temperature drops below $T_c$, given an example of CDM parameters at TeV scale. The gap equation is defined by $\frac{\partial V_{\rm eff}}{\partial \Delta}=0$. Thus, we take the partial derivative of $\frac{1}{4}M^2\Delta^2-\mathcal{L}_1(\Delta, T)$:
\begin{equation}
    \frac{1}{4}M^2 = \int \frac{d^3k}{(2\pi)^3}\left[\frac{\tanh(\frac{1}{2}\beta Q_k^-)}{2Q_k^-}+\frac{\tanh(\frac{1}{2}\beta Q_k^+)}{2Q_k^+}\right]=\int \frac{d^3k}{(2\pi)^3}\left[\frac{1-2f(Q_k^-)}{2Q_k^-}+\frac{1-2f(Q_k^+)}{2Q_k^+}\right]
\end{equation}
where $f(x) = \frac{1}{e^{\beta x}+1}$.

\subsection{Critical Temperature}
By definition, since the gap at the critical temperature gives $\Delta(T_c)=0$, the critical gap equation is:
\begin{equation}
    \frac{1}{4}M^2 = T_c\sum_{\omega_n}\int \frac{d^3k}{(2\pi)^3}\left[\frac{1}{\omega_n^2 + (k-\kappa)^2}+\frac{1}{\omega_n^2 + (k+\kappa)^2}\right]
\end{equation}

After substituting variables and similar dimensional regularization as in Eq.(\ref{eqn:dimreg}):
\begin{eqnarray}
    \frac{\pi^2M^2}{4}&=&T_c\sum_{n}\left[\kappa^2\int_{0}^{\infty}\frac{d\xi}{\xi^2+\omega_n^2}+\int_{0}^{\infty}\frac{\xi^2 d\xi}{\xi^2+\omega_n^2}\right] \cr
    &=&T_c\sum_{n}\left[\frac{\kappa^2}{2}\frac{\pi}{\abs{\omega_n}}+\frac{1}{4\pi}\frac{(2\pi)^{(3-\epsilon)/2}}{\Gamma(3/2-\epsilon/2)}\Gamma\left(\frac{3-\epsilon}{2}\right)\Gamma\left(\frac{\epsilon-1}{2}\right)\frac{\omega_n^{1-\epsilon}}{2\Gamma(1)}\right]\cr
    &=&\pi T_c\sum_{n}\left[\frac{\kappa^2}{2}\frac{1}{\abs{\omega_n}}-\frac{\omega_n}{\sqrt{2}}\right]+O(\epsilon)
\end{eqnarray}
at leading order. We notice that positive and negative contributions to the sum over $\omega_n$ cancel out. Plugging in $\omega_n=T_c(2n+1)\pi$ and a summation cutoff $N=\frac{\Lambda_{UV}}{4\pi T_c}$, we arrive at:
\begin{equation}
    \frac{\pi^2M^2}{4} =\frac{\kappa^2}{2}\ln(\frac{e^{\gamma_e}\Lambda_{UV}}{\pi T_c}).
\end{equation}
Therefore, $T_c=\frac{e^{\gamma_e}\Delta_0}{\pi}$, where $\gamma_e$ is the Euler-Mascheroni constant. In the Appendix, we verify the $T_c$ formula with an alternative method. The termination of this phase transition as $T\rightarrow 0$ is less well-defined, because temperature contributions only relax but do not completely vanish. The following analytical fit has been compared to the numerical solution and proved useful \citep{tinkham2004introduction}:
\begin{equation}
    \Delta \simeq \Delta_0\tanh(\pi e^{-\gamma_e}\sqrt{\frac{T_c}{T}-1}).
\end{equation}
This indicates, for example, that $\Delta \simeq 0.9\Delta_0$ at $T \simeq 0.59 T_c$.

Before embedding this model in a cosmological context, we compute kinetic contributions to the effective field theory at $\mathcal{O}\left(\partial_{\mu}\Delta\right)^2$. 

\section{Kinetic Dispersion}

To compute first-order kinetic fluctuations to the theory, we consider the self-energy diagram below. Since we work with a contact interaction without mediating bosons, we will set $k=0$ in the final step. 

\vspace{-.6cm}
\begin{figure}[H]
    \centering
    \includegraphics[width=0.7\linewidth]{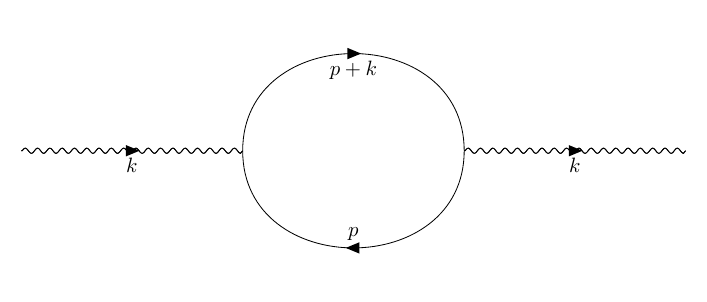}
\end{figure}
\vspace{-1cm}
At finite temperature, the diagram corresponds to an amplitude
\begin{equation}
    \mathcal{M}=-\text{Tr}\left[i \mathcal{G}(p)i \mathcal{G}(p+k)\right]=-\sum_{i\omega_n}\int\frac{d^3p}{(2\pi)^3}\text{Tr}\left[i \mathcal{G}(p)i \mathcal{G}(p+k)\right].
\end{equation}
Since $\mathcal{G}^{-1}(p)=-(\slashed{p}-\Delta+\kappa\gamma^0\gamma^5)$ for our given equation of motion, the fermion propagator is the Green's function $\mathcal{G}(p)=-i (\slashed{p}-\Delta+\kappa\gamma^0\gamma^5)^{-1}$, which simplifies to
\begin{equation}
    \mathcal{G}(p)=-i\frac{(\slashed{p}+\Delta+\kappa\gamma^0\gamma^5)(p^2+\Delta^2+\kappa^2+\kappa\{\slashed{p},\gamma^0\gamma^5\})}{4\kappa^2(p_0^2+p^2)-(p^2+\Delta^2+\kappa^2)^2}.
\end{equation}
Defining $\xi(p)=-4\kappa^2(p_0^2+p^2)+(p^2+\Delta^2+\kappa^2)^2$, the integrand $\mathcal{I}$ takes the form:
\begin{eqnarray}
    \mathcal{I}&=&\frac{4\left(\kappa^6-p^2\left[-\kappa^4+(\Delta^2+q^2)(\Delta^2-p\cdot q)+\kappa^2\left(p_0q_0+\vec{p}\cdot\vec{q}+2\abs{q}^2-q^2\right)\right]\right)}{\xi(p)\xi(q)} \cr
    &+&\frac{4\kappa^4\left(\Delta^2-p_0q_0-2\abs{p}^2+\vec{p}\cdot\vec{q}+q^2-2\abs{q}^2\right)}{\xi(p)\xi(q)}-\frac{4\Delta^2(\Delta^2+q^2)(\Delta^2-p\cdot q)}{\xi(p)\xi(q)} \cr
    &-&\frac{4\kappa^2\left[\Delta^4+(6\Delta^2+4p_0q_0+q^2)(\vec{p}\cdot\vec{q})+2\Delta^2\left(p_0q_0+\abs{q}^2\right)\right]}{\xi(p)\xi(q)} \cr
    &-&\frac{4\kappa^2\left[q^2p_0q_0+2\abs{p}^2\left(\Delta^2+q^2-2\abs{q}^2\right)\right]}{\xi(p)\xi(q)}
\end{eqnarray}
where $p=q+k$, $q^2=-q_0^2+\abs{q}^2$, $p\cdot q=-p_0q_0+\vec{p}\cdot\vec{q}$, and $\vec{p}\cdot\vec{q}=\abs{p}\abs{q}\cos\theta$. The coefficients of the quadratic perturbation $\xi^{\mu\nu}\partial_{\mu}\delta\Delta\partial_{\nu}\delta\Delta$ around $\Delta_0$, where $\delta\Delta=\sigma/\sqrt{\xi^{00}}$, are obtained by taking the $k^2-$derivatives of $\mathcal{M}$ and then integrating over $d^4p$ in the zero-temperature case. Thus:
\begin{equation}\label{finite-deriv}
    \xi^{00}=\left.\frac{d\mathcal{M}}{dk^2_0}\right\vert_{k_0=\abs{k}=0}=\frac{2\left(\Delta^4-(p-\kappa)^4+\omega^4+6\Delta^2\omega^2\right)}{\left(\Delta^2+(p-\kappa)^2-\omega^2\right)^4}+(\kappa\rightarrow -\kappa).
\end{equation}
Integrating over $\frac{d^4p}{(2\pi)^4}$ and using the residue theorem: 
\begin{equation}
    \xi^{00}=\frac{\kappa^2}{12\pi^2\Delta^2}\left[1-\frac{\Delta^2}{\kappa^2}\left(4+3\ln(\frac{\Delta}{\Lambda_p})\right)\right]
\end{equation}
where $\Lambda_p$ is the ultraviolet cutoff of the integral. Similarly, we apply the residue theorem to the $\abs{k}^2-$ derivative:
\begin{equation}
    \xi^{33}=\left.\frac{d\mathcal{M}}{d
    \abs{k}^2}\right\vert_{k_0=\abs{k}=0}=-\frac{\kappa^2}{12\pi^2\Delta^2}\left[\frac{1}{3}+\frac{\Delta^2}{\kappa^2}\left(3\ln(\frac{\Lambda_p}{\Delta})-5\right)\right].
\end{equation}
Since around $\Delta_0$:
\begin{equation}
    V(\Delta)\simeq V(\Delta_0)+\frac{\kappa^2}{\pi^2}(\Delta-\Delta_0)^2=V(\Delta_0)+\frac{\kappa^2}{\pi^2}\frac{\sigma^2}{\xi^{00}}
\end{equation}
we gain the leading-order dispersion relationship at zero temperature: 
\begin{equation}\label{eqn:dispersion}
    \omega^2\simeq\frac{\xi_{33}}{\xi^{00}}k^2+\frac{\kappa^2}{\pi^2\xi^{00}}=\frac{1}{3}k^2+m_{\rm eff}^2=\frac{1}{3}k^2+12\Delta_0^2.
\end{equation}
This gives a negligible LSS group velocity of:
\begin{equation}
    v_g=\frac{d\omega}{dk}=\frac{1/\sqrt{3}}{\sqrt{1+m_{\rm eff}^2/c_p^2k^2}}\simeq \frac{k}{m_{\rm eff}}\sim\mathcal{O}(10^{-36})
\end{equation}
At finite temperature, we perform a Matsubara sum over fermionic frequencies for the expression in Eq.(\ref{finite-deriv}):
\begin{equation}\label{eqn:matsubara}
    \frac{1}{\beta}\sum_{n=-\infty}^{\infty}\frac{2\left(\Delta^4-(p-\kappa)^4+\omega_n^4-6\Delta^2\omega_n^2\right)}{\left(\Delta^2+(p-\kappa)^2+\omega_n^2\right)^4}+\frac{1}{\beta}\left(\kappa\rightarrow -\kappa\right).
\end{equation}
\begin{figure}
    \centering
    \includegraphics[width=0.7\linewidth]{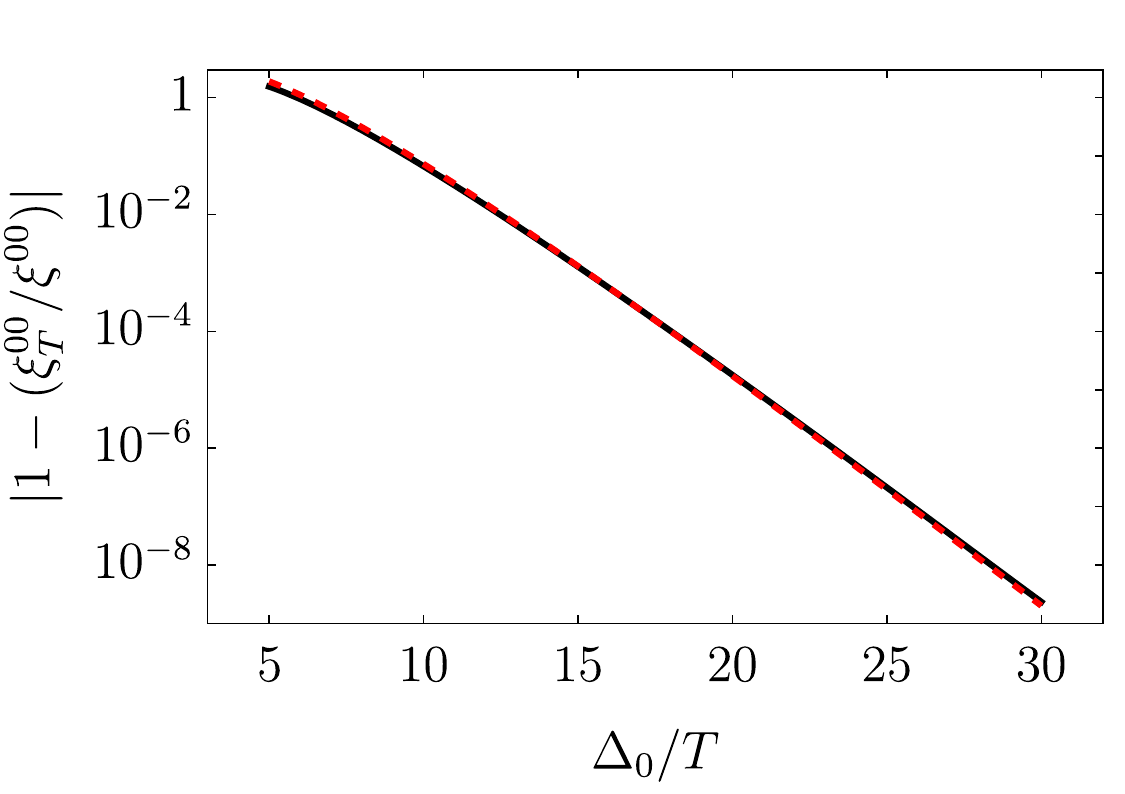}
    \caption{Finite temperature corrections to the zero temperature $\xi_{00}$. Corrections are exponentially suppressed for large $\Delta_0/T$. Fitted by $e^{-x}x^2\ln(x)$, where $x=\beta\Delta$.}
    \label{fig:thermalcorrection}
\end{figure}
The full result is too long to be informative, hence given in the appendix. But for $T\lesssim 0.1\Delta_0$ or $w\lesssim 0.1$, the summation result simplifies to:
\begin{align}
    \frac{e^{4\beta\sqrt{\Delta^2+(p-\kappa)^2}}}{4\left(1+e^{\beta\sqrt{\Delta^2+(p-\kappa)^2}}\right)^4}\frac{(p-\kappa)^2}{\left(\Delta^2+(p-\kappa)^2\right)^{5/2}}+(\kappa\rightarrow -\kappa).
\end{align}
This period is the most relevant for investigating the candidate's coldness. The subsequent momentum integral $\frac{d^3p}{(2\pi)^3}$ must be evaluated numerically. Using the same TeV-scale parameters as in Fig.(\ref{fig:phase-transition}) and Fig.(\ref{fig:rho}), the finite temperature corrections to $\xi^{00},\, \xi^{33}$ are found to be exponentially suppressed. For example, the thermal corrections to $\xi^{00}$ are well-described by $\abs{1 - \xi^{00}_T/\xi^{00}} \simeq (x \ln x)^2 e^{-x}$ for $x = \beta\Delta$ in the range $x=5 - 30$; for larger values of $\beta\Delta$ the exponential suppression dominates, Fig.(\ref{fig:thermalcorrection}).  

\section{Size of Cooper Pairs}

Fermion pairs are spaced apart at the correlation length scale. The fermion-antifermion correlation function $\mathcal{G}(x,t;y,t')=\langle \bar\psi(x,t)\psi(y,t')\rangle$ describes the probability for a particle created at $(x,t)$ to be annihilated with its antiparticle at $(y,t')$. Thus, the length scale associated with the Green's function captures the characteristic size of fermionic Cooper pairs. $\mathcal{G}(x,t;y,t')$ is obtained via a Fourier transform of the Green's function $\mathcal{G}(\vec{p},\omega)$:
\begin{equation}
    \mathcal{G}(\vec{x},t;\vec{y},t'))=\int \frac{d^3p}{(2\pi)^3}\frac{1}{\beta}\sum_{\omega_n}\mathcal{G}(\vec{p},\omega_n)e^{i\vec{p}\cdot(\vec{x}-\vec{y})-i\omega(t-t')}
\end{equation}
where we assumed finite temperature. First recall that
\begin{equation}
    \mathcal{G}(\vec{p},\omega)=-i\frac{(\slashed{p}+\Delta+\kappa\gamma^0\gamma^5)(p^2+\Delta^2+\kappa^2+\kappa\{\slashed{p},\gamma^0\gamma^5\})}{4\kappa^2(\omega^2+p^2)-(p^2+\Delta^2+\kappa^2)^2}.
\end{equation}
Taking its trace, we get
\begin{equation}
    \Tr(i\mathcal{G}(\vec{p},\omega))=\frac{4\Delta(p^2+\Delta^2+\kappa^2)}{-(p^2+\Delta^2+\kappa^2)^2+4\kappa^2(p^2+\omega^2)}.
\end{equation}
Here, we are interested in the spatial correlation function with $r=\abs{\vec{x}-\vec{y}}$ and same time $t=t'$. Therefore,
\begin{equation}
    \xi(r)=\Tr(i\mathcal{G}(\vec{x},t;\vec{y},t))=\int \frac{d^3p}{(2\pi)^3}\frac{1}{\beta}\sum_{\omega_n}\Tr(i\mathcal{G}(\vec{p},\omega_n))e^{i\vec{p}\cdot(\vec{x}-\vec{y})}.
\end{equation}
In evaluating the Matsubara sum over fermionic frequencies, we used:
\begin{equation}
    \frac{1}{\beta}\sum_{i\omega_n}\frac{1}{(i\omega_n)^2-a^2}=-\frac{1}{2a}\left(1-\frac{2}{1+e^{\beta a}}\right)
\end{equation}
for $a\in\mathbb{R}$. We arrive at the integral
\begin{equation}
    \xi(r) = -\int \frac{d^3 p}{(2 \pi)^3} e^{i \vec p \cdot(\vec x- \vec y)}\frac{\Delta}{\kappa} \frac{p^2 + \Sigma^2}{\sqrt{(p^2 + \Sigma^2)^2 - 4 \kappa^2 p^2}} \tanh\left[\frac{\beta}{4 \kappa} \sqrt{(p^2 + \Sigma^2)^2 - 4 \kappa^2 p^2} \right]
\end{equation}
where $\Sigma^2 = \kappa^2 + \Delta^2$. Here, the argument of $\tanh$ is positive, with a minimum value $\frac{1}{2} \beta \Delta$. For $\Delta/T\gg 1$, as is the case for our CDM model, the $\tanh$ function is effectively one. Thus,
\begin{equation}
    \xi(r) \simeq -\int \frac{d^3 p}{(2 \pi)^3} e^{i \vec p \cdot(\vec x- \vec y)}\frac{\Delta}{\kappa} \frac{p^2 + \Sigma^2}{\sqrt{(p^2 + \Sigma^2)^2 - 4 \kappa^2 p^2}}.
\end{equation}
We now define $f(p) = (p^2 + \Sigma^2)/\sqrt{(p^2 + \Sigma^2)^2 - 4 \kappa^2 p^2}$ and observe that $f\to 1$ at $p\to 0,\,\infty$. However, $f$ is sharply peaked at $p = \kappa$ with amplitude $\kappa/\Delta$. Numerical simulations reveal its proximity to a Gaussian function with narrow width $\sigma = \sqrt{3} \Delta$: 
\begin{equation}
    f \simeq 1 + \frac{\kappa}{\Delta}\exp\left({-\frac{1}{2}\frac{(p-\kappa)^2}{\sigma^2}}\right).
\end{equation}
Using the above approximation for $f$, the correlation function is now
\begin{equation}
    \xi(r) \simeq -\frac{\Delta}{\kappa} \delta(\vec x - \vec y) - \frac{1}{2 \pi^2} \int_0^\infty dp\, p^2 \, \frac{\sin(p r)}{p r}\, \exp\left({-\frac{1}{2}\frac{(p-\kappa)^2}{\sigma^2}}\right).
\end{equation}
Since $\sigma \ll \kappa$, the Gaussian behaves as a delta function,
\begin{equation}
    \frac{1}{\sqrt{2 \pi \sigma^2}}\exp\left({-\frac{1}{2}\frac{(p-\kappa)^2}{\sigma^2}}\right) \simeq \delta(p-\kappa).
\end{equation}
In this case, the correlation function evaluates to
\begin{equation}
    \xi(r) \simeq -\frac{\Delta}{\kappa} \delta(\vec x - \vec y) - \frac{\sqrt{6 \pi}}{2 \pi^2} \Delta \kappa^2 \frac{\sin(\kappa r)}{\kappa r}.
\end{equation}
For the spatially separated fermions, the Dirac delta function is zero. We observe that the characteristic length scale on which the correlation function decays is $r_c \sim \kappa^{-1}$. In our example of $\mathcal{O}(\kappa)$ at TeV-scale, the correlation length is microscopic. This shows that on astrophysical scales concerned with dark matter, condensed fermion pairs behave effectively as one particle. 


\chapter{Cosmological Implications}\label{dmde}
\section{Thermal History}
In this chapter, we follow the standard definitions of pressure and energy density, $p=-V$ and $\rho=T\frac{dp}{dT}-p$ \citep{Kolb:1981hk}. This definition takes into account all fermions in the grand canonical ensemble with interaction specified by Eq.(\ref{eqn:dirac}). As a result of the fermion thermal field theory, there are three distinct epochs in their cosmic evolution. 

\subsection{Radiation Epoch}
At high temperatures, $T \gg \kappa, m$, the gap field lies at the potential minimum $\Delta = 0$. Then by definition above and Eq.(\ref{eqn:HighTVeff}):
\begin{equation}
    \rho = \frac{7}{8} \frac{g \pi^2}{30}T^4 = 3p.
\end{equation}
This agrees exactly with the energy density and pressure of a relativistic gas of fermions with degeneracy factor $g=4$, which includes both particle and antiparticle and two helicities. Furthermore, the fermion equation of state $w=\frac{p}{\rho}=\frac{1}{3}$ behaves like standard cosmological radiation. Thus, we assume these fermions were once in thermal equilibrium with the SM, sharing the same temperature $T_{\Delta}=T_{\gamma}$ as photons. We further assume scattering processes maintaining equilibrium later became inefficient and the fermion system decoupled, so that $T_\Delta$ evolved separately, adiabatically thereafter: $s_\Delta=(\rho_\Delta+p_\Delta)/T_\Delta \propto a^{-3}$, where $a$ is the cosmic expansion scale factor. The subsequent dropout of SM species from equilibrium heats the photons so that $T_\gamma > T_\Delta$. The presence of this additional, relic radiation contributes to the rate of cosmic expansion, which in turn is tightly constrained through observations of the light element abundances predicted by Big Bang Nucleosynthesis (BBN), and through observations of the cosmic microwave background (CMB). These constraints can be satisfied by enforcing a sufficiently early time for decoupling. For example, decoupling prior to the QCD phase transition would be adequate to lower $T_\Delta$ relative to $T_\gamma$ to satisfy the Planck 2018 bound $\Delta N_{\rm eff} < 0.3$ (95\% CL, one-tailed Planck TT,TE,EE+lowE+lensing+BAO \citep{Planck:2018vyg}) on the effective number of additional, relativistic particle species, as customarily measured in units of massless neutrinos. This constraint is easily satisfied by, for example, the TeV parameters in Fig.(\ref{fig:phase-transition}). 

\subsection{"Rapid-Decay" Epoch}
As the temperature drops below $\kappa$, and the gap remains at $\Delta = 0$, we see that the effective potential is dominated by the $I_0$ term in Eq.~(\ref{eqn:vt}), where $I_0(0)=\pi^2/12$. The resulting energy density and pressure are $p_\Delta = \rho_\Delta = \kappa^2 T^2/6$ with $w_\Delta=1$. For this equation of state, adiabatic evolution yields $\rho_\Delta \propto (1+z)^6$, where $z$ is the expansion redshift. Hence, the $\Delta$ field decays faster than both radiation and non-relativistic matter. This $\kappa$-dominated epoch plays a comparable role to freeze out for thermal WIMPs, dropping the field out of equipartition with the cosmic fluid, and setting the relic density for the subsequent matter-dominated evolution.

\begin{figure}
    \centering
    \includegraphics[width=0.75\linewidth]{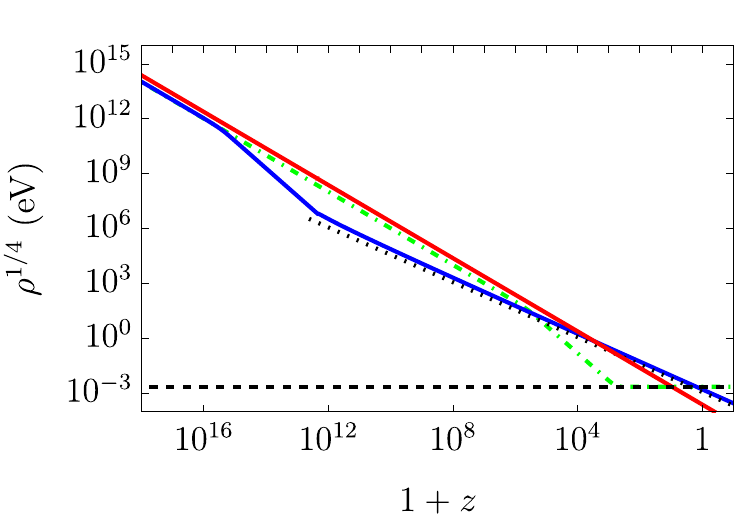}
    \caption{Cosmic energy density $\rho^{1/4}$ vs. redshift, $z$. Standard Model radiation (red), $\Delta$ field with $\Omega_{\rm CDM}=0.25$ (blue), cosmological constant dark energy with $\Omega_{\rm DE}=0.7$ (black dashed), and baryons with $\Omega_{\rm B}=0.05$ (black dotted). Same parameters for CDM are used as in Fig.(\ref{fig:phase-transition}). The proposed, metastable dark energy model is also shown (green, dot-dashed).}
    \label{fig:rho}
\end{figure}

\subsection{Below Critical Temperature}
\begin{figure}
    \centering
    \includegraphics[width=0.75\linewidth]{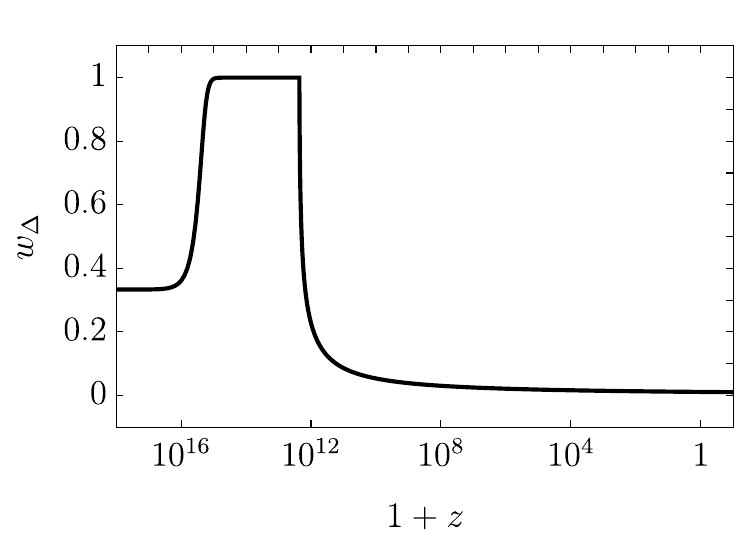}
    \caption{Equation of state $w_\Delta$ versus redshift, $z$. There are four distinct eras: Radiation, $w_\Delta=1/3$; $\kappa$-domination, $w_\Delta = 1$; Phase Transition, at $z\sim 10^{12}$; Pressureless matter, at $z\lesssim 10^{12}$. Same parameters are used as in Fig.\ref{fig:phase-transition}.}
    \label{fig:w}
\end{figure}

The $\kappa$-era ends when the temperature reaches $T_c$. For the massless case, the field undergoes a second-order phase transition, where the minimum smoothly transfers from $\Delta=0$ to $\Delta_0$ as $T \to 0$. To model this behavior, the integrals in Eq.~(\ref{eqn:vt}) must be evaluated numerically. However, for $T \lesssim T_c/2$, we observe $w_\Delta \simeq T/\Delta_0$. As $T$ drops well below the gap field strength, the condensate approaches the state of ideal, pressureless matter. Given the Friedmann Equation and entropy conservation in cosmology (with density $s=\frac{\rho+p}{T}$), 
\begin{eqnarray}
    \frac{d\rho}{da} &=& - \frac{3}{a}\rho(1 + w) \\
    \frac{\rho(1+w)}{T} a^3 &=& \frac{\rho_0(1+w_0)}{T_0} a_0^3.
\end{eqnarray}
we can solve for $\frac{da}{d\rho}$:
\begin{equation}
    \left(\frac{a}{a_0}\right)^3 = \frac{\rho_0}{\rho}\left(1 + \frac{T_0}{T_0 + \Delta_0}\log\frac{\rho_0}{\rho}\right)^{-1}
\end{equation}
The equation of state in terms of energy density is
\begin{equation}
    w = \left(\frac{\Delta_0}{T_0} + \log \frac{\rho_0}{\rho}\right)^{-1},
\end{equation}
suggesting an analytic solution
\begin{equation}
    (1+z)^3 = \frac{w_0 (1+w_\Delta)}{w_\Delta(1+w_0)}\, e^{\tfrac{1}{w_0}-\tfrac{1}{w_\Delta}},
    \label{eqn:w}
\end{equation}
where $w_0=T_0/\Delta_0$ is the present-day value of $w_\Delta$. For the example model, $w_0\simeq 0.01$ and $w_\Delta \lesssim 0.02$ at equality. Recent studies of the dark-matter equation of state \citep{Kopp:2018zxp,Ilic:2020onu} based on current CMB and large scale structure (LSS) data such as Planck \citep{Planck:2018vyg} and the Sloan Digital Sky Survey \citep{BOSS:2013rlg} find a sub-percent level constraint on $w$ near recombination at the $99\%$ confidence level. However, we caution that the studies assumed $w$ in bins joined by sharp transitions, in contrast with the slow, continuous evolution predicted in  Eq.~(\ref{eqn:w}). The equation-of-state history and energy density in our model are illustrated in Figs.(\ref{fig:rho}-\ref{fig:w}).

\section{Coldness of Dark Matter}

By Eq.(\ref{eqn:dispersion}), the dispersion relationship in expanding spacetime is $\omega^2=\frac{1}{3}k^2+12a^2\Delta_0^2$ at $T=0$. At finite temperature, we see from Fig.(\ref{fig:thermalcorrection}) that corrections to $\xi_{00}$ and $\xi_{33}$ are exponentially suppressed for $T\lesssim 0.1\Delta_0$ or $w\lesssim 0.1$. Furthermore, this run of $\beta\Delta$ covers the epoch $z \lesssim 10^{12}$, relevant for the CMB and LSS. The physically relevant group velocity is therefore $v_g = \frac{d\omega}{dk} \sim {\cal O}\left(\frac{k}{a \Delta_0}\right)$, which is negligible on cosmological scales. For the example model of Fig.(\ref{fig:phase-transition}), $\Delta_0 \sim 100$~eV whereas $k/a\sim 10^{-33}$~eV for the present-day Hubble radius. 


The Lyman-$\alpha$ spectrum provides a critical upper bound on the comoving free-streaming length $\chi_{\rm FS}$ of CDM candidates. Since $v$ is negligible after becoming non-relativistic:
\begin{equation}
    \chi_{\rm FS}=\int vd\tau\simeq \int_{\rm rel} \frac{vdt}{a}=\int_0^{10^{-12}} \frac{da}{a^2H(a)}
\end{equation}
where $\tau$ is proper time. The candidate remains relativistic until $z=10^{12}$ (Fig.\ref{fig:w}) and the group velocity $v$ becomes negligible in the non-relativistic phase. Assuming non-radiation energy contributions are negligible in the relativistic phase:
\begin{equation}
    H(a)\simeq H_0\sqrt{\Omega_{\rm r,0}}\left(\frac{a_0}{a}\right)^2.
\end{equation}
The Hubble distance $H_0\simeq 3000 h^{-1}$Mpc, $\Omega_{\rm r,0}\simeq 4.3\times 10^{-5}$, $h\sim 0.7$. Therefore, $\chi_{\rm FS}\simeq 6.5\times 10^{-7}$Mpc, far below the $1$Mpc length scale at which Lyman-$\alpha$ and CMB constraints are capable of distinguishing between cold and warm dark matter.

\section{Chiral Asymmetry Parameter}
In this initial foray, we have not attempted to embed our low energy model in a realistic particle theory. However the axial asymmetry, necessary for the existence of the nontrivial gap solution, suggests a possible connection to leptogenesis models that explain the matter - antimatter asymmetry of the Universe \citep{Davidson:2008bu}. Around $\Delta_0$, the thermal ensemble average of the axial number density $n_A=\langle \bar\psi \gamma^0 \gamma^5 \psi \rangle$ and the overall fermion number density $n_{\Delta}=\langle \bar\psi \psi \rangle$ are given by:
\begin{eqnarray}
    n_A&=&-\frac{\partial V(\Delta_0,T)}{\partial \kappa}=\frac{\kappa\Delta_0^2}{2\pi^2}\left(1+\frac{\pi^2M^2}{\kappa^2}\right)+\frac{\kappa T^2}{3} \cr
    n_{\Delta} &=&\left.\frac{\partial V(\Delta,T)}{\partial \Delta}\right\vert_{\Delta_0}=\frac{1}{2}M^2\Delta_0.
\end{eqnarray}
To quantify the degree of asymmetry, we define a chiral asymmetry parameter $r_A \equiv n_A / n_\Delta$. By the present day, with $T \ll \Delta_0$, $r_A=\frac{\Delta_0}{\kappa}\left(1+{\kappa^2}/{\pi^2 M^2}\right)$. The exponential suppression of $\Delta_0$ easily facilitates parameters such that $r_A\simeq 6.5\times 10^{-10}$, suggestive of the baryon asymmetry parameter $\eta = n_B/n_\gamma = 6.10(\pm 0.04) \times 10^{-10}$ \citep{Planck:2015fie}. Furthermore, this suggests a renormalization scheme. In principle, given a leptogenesis scheme that connects $r_A$ to $\eta$, then if $M$ and $\kappa$ are fixed, one can use the observed baryon-to-photon ratio to determine the cutoff, $\Lambda_{UV}$.

\section{Dark Energy}
\begin{figure}[H]
    \centering
    \includegraphics[width=0.8\linewidth]{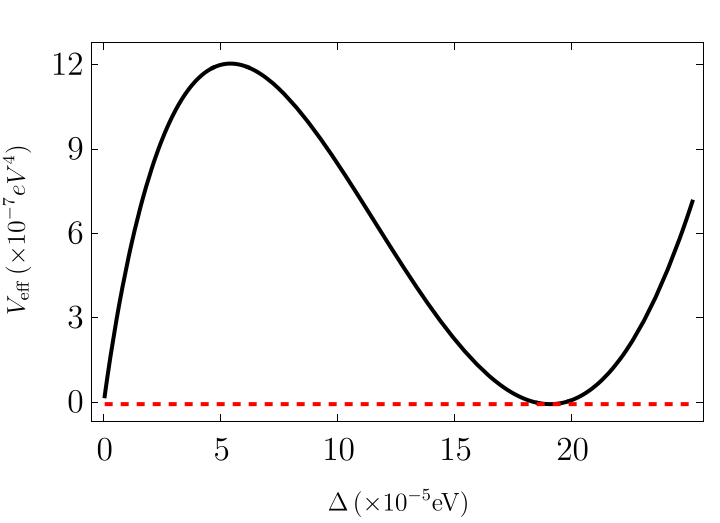}
    \caption{Effective Potential in the Massive Case. As before, $V(\Delta_{\rm min})=0$. Notice that $V(0)-V(\Delta_{\rm min})$ is tiny compared to the barrier height $10^{-6}$eV$^4$ and width $\sim 10^{-4}$eV, exponentially suppressing the tunneling probability to $\Delta_{\rm min}$.}
    \label{fig:massiveV}
\end{figure}

We consider the massive NJL fermion condensate as a candidate for cosmological constant-like dark energy. We restrict attention to the case $m \ll \Delta_0$, whereby the thermal history of massive NJL fermions is similar to the massless case during the periods $w_\Delta=1/3$ and $w_\Delta=1$. However, the $\tfrac{1}{4}M^2(\Delta+m)^2$ term in Eq.~(\ref{eqn:vloop}) has positive slope at the origin, resulting in a potential barrier that traps the field at $\Delta=0$ at low temperatures. As before, we shift the potential such that $V(\Delta_m)=0$, where $\Delta_m\simeq\Delta_0 + m \ln(\Delta_0/\Lambda_{UV})$ is the global minimum. Instead of undergoing a phase transition at $T_c$, the field remains in a potential-dominated metastable state with equation of state $w_\Delta \to -1$. For an example model, we use $(M,\,\Lambda_{UV},\,\kappa,\,m) \sim$ $(128,\, 96,\, 80,\, 7.5\times 10^{-6})$~eV to obtain $\Omega_{\rm DE} \simeq 0.7$. This metastable state is separated from the true vacuum by a barrier $\sim 10^{-6}$~eV${}^4$ high and $\sim 10^{-4}$~eV wide as shown in Fig.(\ref{fig:massiveV}). 

To estimate the lifetime of this state, we consider the bubble nucleation rate $\Gamma = A\exp(-S_E)$, where $S_E$ is the Euclidean action of the tunneling path joining the metastable and true vacua. We start with:
\begin{equation}
    S_E=\int dt_Ed^3x \left[\frac{1}{2}\left(\frac{d\Delta}{dt_E}\right)^2+\frac{1}{2}\left(\partial_i\Delta\right)^2+V(\Delta)\right]
\end{equation}
where $t_E=it$ is Euclidean time. We now assume $O(4)$ symmetry in 4-Euclidean flat space. Defining $r^2=t_E^2+\abs{x}^2$, we can rewrite the action with $\Delta(r)$:
\begin{equation}\label{euclid-action}
    S_E=2\pi^2\int dr r^3 \left[\frac{1}{2}(\partial_r\Delta)^2+V(\Delta)\right]
\end{equation}
since the three sphere has volume $2\pi^2r^3$. The corresponding equation of motion is gained from the variational principle $\frac{\delta S_E}{\delta\Delta}=0$. We recognize that
\begin{equation}
    \frac{1}{2}r^3(\partial_r\Delta)^2=\frac{1}{2}\left[\partial_r(r^3\Delta\partial_r\Delta)-3r^2\Delta\partial_r\Delta-r^3\Delta\partial_r^2\Delta\right].
\end{equation}
Since the total derivative vanishes under integration,
\begin{equation}\label{eqn:Delta-eom}
    \frac{3}{r}\frac{d\Delta}{dr}+\frac{d^2\Delta}{dr^2}-\frac{dV}{d\Delta}=0. 
\end{equation}
For the chosen parameters, we are justified to work in the thin-wall limit because the phase transition happens over a small interval of $r$. Since time elapsed is large ($r$ large) before $\Delta$ begins rolling \citep{Kolb:1981hk}, $\frac{1}{r}$ is negligible and Eq.(\ref{eqn:Delta-eom}) becomes:
\begin{equation}
    \frac{d^2\Delta}{dr^2}=\frac{dV}{d\Delta}.
\end{equation}
Multiplying both sides by $\frac{d\Delta}{dr}$,
\begin{eqnarray}
    \frac{d\Delta}{dr}\frac{d^2\Delta}{dr^2}&=&\frac{dV}{d\Delta}\frac{d\Delta}{dr}, \cr
    \frac{1}{2}\frac{d}{dr}\left(\frac{d\Delta}{dr}\right)^2&=&\frac{dV}{dr},\cr
    \frac{d\Delta}{dr}&=&\pm\sqrt{2V(\Delta)}.
\end{eqnarray}
Therefore, we get $r=\int_{\Delta}\frac{d\Delta'}{\sqrt{2V(\Delta')}}$. Substituting back into Eq.(\ref{euclid-action}), we find the bubble radius that minimizes the Euclidean action. $S_E$ can now be approximated:
\begin{equation}
S_E=\frac{27\pi^2\left[\int_{0}^{\Delta_m}d\Delta\sqrt{2 V(\Delta)}\right]^4}{2\left(V(0)-V(\Delta_m)\right)^3}\simeq 2.8\times 10^7.
\end{equation}
For any reasonable value of $A$, a first-order phase transition to $\Delta_m$ is exponentially suppressed, and the time to tunnel is much longer than the age of the Universe.

For this model, the $\kappa$-era begins when the SM photon temperature reaches $T_\gamma \sim 1$~keV, near $z\sim 10^6$. Decoupling of the fermions from the SM must occur sufficiently early to satisfy BBN constraints on the excess degrees of freedom of the cosmic fluid. The $\kappa$-era ends and the field is potential dominated from $z\sim 10^3$ onwards. This model also addresses various problems that plague many dark energy models \citep{Velten:2014nra}. At early times the fermions are in thermal equilibrium with the SM, so no fine tuning of initial conditions is required. The exponential suppression of $\Delta_m$, and the requirement that $m \ll \Delta_m$, necessary for the existence of a minimum, helps explain the smallness of dark energy. And the chemical potential, which separates CDM and dark energy from the thermal bath, helps explain the coincidence of energy densities. 

\section{Conclusions}

In this thesis, we present a novel candidate for CDM and dark energy. In analogy with superconductivity, the dark matter particles are non-relativistic Cooper pairs of massless fermions. For dark energy, the potential energy is associated with pairs of massive fermions. We highlight the important role of the thermal history, in particular the mechanism of the chemical potential, in setting the relic abundance. The scenario further benefits from the exponentially suppressed physical scales, well below the cutoff of this low energy theory, that are generated by the gap field solution. The exponential suppression may help explain two small numbers in cosmology: the smallness of dark energy density $\left(2\times 10^{-3}\text{eV}\right)^4$, and the axial asymmetry $\eta\sim 10^{-10}$ which may link dark matter to the baryon asymmetry. A unique prediction of the model is a non-zero, time-evolving equation of state, $w_\Delta$. Data from CMB and LSS surveys may be able to constrain $w_\Delta$ down to the percent level across certain redshift bands \citep{Kopp:2018zxp,Ilic:2020onu}, and thereby test the energy scales and interactions of the fermion system. More precise constraints on $\Delta N_{\rm eff}$ expected from future CMB missions such as the Simons Observatory \citep{SimonsObservatory:2018koc} and CMB-S4 \citep{CMB-S4:2016ple} will further inform the energy scales and decoupling of these species. The slightly faster decay of CDM will also shift the geometry of the CMB acoustic oscillations, and may help relieve another outstanding problem of cosmology, the Hubble Tension \citep{Freedman:2017yms}. We reserve for future work to investigate the imprint on the CMB and LSS, to consider the collisional nature of this species of dark matter, and embed this low energy theory in a more complete model of particle physics.

\backmatter




\appendix
\newpage
\renewcommand{\thechapter}{}
\renewcommand{\thesection}{\Alph{section}}
\chapter{Appendix}

\section{Finite Temperature Matsubara Summation}

Below, we attach the full result for the Matsubara summation in Eq.(\ref{eqn:matsubara}):
\[
\frac{
\begin{aligned}
    &\frac{1}{\beta}\sum_{n=-\infty}^{\infty}\frac{2\left(\Delta^4-(p-\kappa)^4+\omega_n^4-6\Delta^2\omega_n^2\right)}{\left(\Delta^2+(p-\kappa)^2+\omega_n^2\right)^4}+\frac{1}{\beta}\left(\kappa\rightarrow -\kappa\right) \\
     =& (\kappa \rightarrow -\kappa) + 3 T^3 (p + \kappa)^2 - 3 e^{\frac{4 \sqrt{\Delta^2 + (p+\kappa)^2}}{T}} T^3 (p+\kappa)^2 \\
     + & 4 e^{\frac{2 \sqrt{\Delta^2 + (p+\kappa)^2}}{T}} \sqrt{\Delta^2 + (p+\kappa)^2} 
    \Big(4 \Delta^4 + p^2 (3 T^2 + 4 \Delta^2) + (3 T^2 + 4 \Delta^2) \kappa^2 + p(6 T^2 \kappa + 8 \Delta^2 \kappa) \Big) \\
    + & e^{\frac{3 \sqrt{\Delta^2 + (p+\kappa)^2}}{T}} 
    \Big(6 p^4 T + 24 p^3 T \kappa - 6 T^3 \kappa^2 + 6 T \kappa^2 (\Delta^2 + \kappa^2) 
    + 6 T^2 \kappa^2 \sqrt{\Delta^2 + (p+\kappa)^2} \\
    - & 4 \Delta^2 (\Delta^2 + \kappa^2) \sqrt{\Delta^2 + (p+\kappa)^2} + p^2 \bigg(-6 T^3 + 6 T \left(\Delta^2 + 6 \kappa^2\right) 
    + 6 T^2 \sqrt{\Delta^2 + (p + \kappa)^2} \\
    - & 4 \Delta^2 \sqrt{\Delta^2 + (p + \kappa)^2}\bigg) + 4 p \kappa \bigg(-3 T^3 + 3 T \left(\Delta^2 + 2 \kappa^2\right) 
    + 3 T^2 \sqrt{\Delta^2 + (p + \kappa)^2} 
    - 2 \Delta^2 \sqrt{\Delta^2 + (p + \kappa)^2}\bigg)\bigg) \\
    - & 2 e^ {\sqrt{\frac{\Delta^2 + (p + \kappa)^2}{T}}}
    \bigg(3 p^4 T + 12 p^3 T \kappa - 3 T^3 \kappa^2 + 3T\kappa^2(\Delta^2+\kappa^2)
    - 3 T^2 \kappa^2 \sqrt{\Delta^2 + (p + \kappa)^2} \\
    + & 2 \Delta^2 \left(\Delta^2 + \kappa^2\right) \sqrt{\Delta^2 + (p + \kappa)^2} + 2 p \kappa \bigg(-3 T^3 + 3 T \left(\Delta^2 + 2 \kappa^2\right) 
    - 3 T^2 \sqrt{\Delta^2 + (p + \kappa)^2} \\
    + & 2 \Delta^2 \sqrt{\Delta^2 + (p + \kappa)^2}\bigg) + p^2 \bigg(-3 T^3 + 3 T \left(\Delta^2 + 6 \kappa^2\right) 
    - 3 T^2 \sqrt{\Delta^2 + (p + \kappa)^2} 
    + 2 \Delta^2 \sqrt{\Delta^2 + (p + \kappa)^2}\bigg)\bigg)
\end{aligned}
}{
12 \big(1 + e^{\frac{\sqrt{\Delta^2 + (p+\kappa)^2}}{T}}\big)^4 T^3 (\Delta^2 + (p+\kappa)^2)^{5/2}
}
\] 

We note that below $T_c=\frac{e^{\gamma_e}}{\pi}\Delta_0$, as $T$ decreases and $\Delta\rightarrow \Delta_0$, the $\left(1 + e^{\frac{\sqrt{\Delta^2 + (p+\kappa)^2}}{T}}\right)^4$ exponential factor in the denominator begins to dominate over every factor except for $e^{4\frac{\sqrt{\Delta^2 + (p+\kappa)^2}}{T}}$. 

\section{Critical Temperature}
We mathematically verify $T_c=e^{\gamma_e}\Delta_0/\pi$. Recall that:
\begin{equation}
    I_n(x) = \int_0^\infty du\, u^n \, \ln{\left(1 + e^{-\sqrt{x^2+u^2}}\right)}.
\end{equation}
Therefore, $I_2'(x) \simeq - x \pi^2 /12 + {\cal O}(x^2)$ for $x \ll 1$. In the $n=0$ case,
\begin{equation}
    I_0'(x) \simeq \frac{1}{2} x \ln x - c x + {\cal O}(x^2),
\end{equation}
for $x \ll 1$. To derive the factor $c$, we introduce a change of variable $u= x \sqrt{z^2-1}$ whereby
\begin{equation}
    I_0(x) = x \int_1^\infty dz\, \frac{z}{\sqrt{z^2-1}}\ln(1 + e^{-x z}).
\end{equation}
Next, expand the log term
\begin{equation}
    \ln(1 + e^{-x z}) = \sum_{n=1}^\infty \frac{(-1)^{n+1}}{n} e^{-n x z}
\end{equation} 
and perform the $z-$integration
\begin{equation}
    I_0(x) = x \sum_{n=1}^\infty \frac{(-1)^{n+1}}{n} K_1(n x)
\end{equation}
where $K$ is a modified Bessel function. Next, differentiate with respect to x to obtain
\begin{equation}
    I_0'(x) = \sum_{n=1}^\infty (-1)^n x K_0(n x).
\end{equation}
Expanding this sum for $x \ll 1$ and keeping only leading terms:
\begin{equation}
    I_0'(x) = -\sum_{n=1}^\infty (-1)^n (\gamma_e + \ln(n x/2))x + {\cal O}{(x^3)}.
\end{equation}
We now insert a factor $Y^{-n}$ into the sum, and then set $Y$ to $1$:
\begin{eqnarray}
    I_0'(x) = -\sum_{n=1}^\infty \frac{(-1)^n}{Y^n} \left.\left[\gamma_e + \ln(n x/2)\right]x \right|_{Y=1}=  \frac{1}{2} x \left(\gamma_e + \ln \frac{x}{\pi}\right).
\end{eqnarray}
Therefore, $c=\frac{\ln\pi-\gamma_e}{2}$. Since $\frac{\partial V_{\rm eff}(\Delta, T_c)}{\partial \Delta} = 0$ to leading order for $0\lesssim\Delta \ll T_c$, we obtain $T_c=\frac{e^{\gamma_e}}{\pi} \Delta_0$.

\section{Helicity Flipping Rate}
In Section 1.2, we assumed conservation of fermion helicity for the given interaction, i.e. helicity flipping rate is zero. Here, we justify this assumption.

The NJL scalar interaction $(\bar\psi\psi)^2/M^2$ is the low-energy EFT of a fermion-antifermion scattering interaction mediated by a scalar boson. To examine the behavior of helicity under this interaction, $L_{\rm int}= -g\phi\bar\psi\psi$, we consider the transition amplitude $|\langle f|S|i\rangle|^2$, where $|f\rangle=|R\rangle$ and $|i\rangle=|L\rangle$ are the corresponding helicity eigenstates. At first order, $S=1+(-ig)\int d^4x\bar\psi(x)\phi(x)\psi(x)$ and the resulting transition amplitude is proportional to $\bar{u}(p') u(p)$, where $u(p)$ is equivalent to the spinor field $\psi(p)$. For notation, we follow the textbook by \citep{Rubbia:2022hry}. Starting from Eq.(11.9),
\begin{equation}\label{eq:dirac_bilinear}
\begin{aligned}
\bar{u}(p') u(p) &= u^\dagger(p') \gamma^0 u(p) \\
&= \begin{bmatrix} u_1^* & u_2^* & u_3^* & u_4^* \end{bmatrix}
\begin{bmatrix} 0 & 0 & 1 & 0 \\ 0 & 0 & 0 & 1 \\ 1 & 0 & 0 & 0 \\ 0 & 1 & 0 & 0 \end{bmatrix}
\begin{bmatrix} u_1 \\ u_2 \\ u_3 \\ u_4 \end{bmatrix} \\
&= u_1^*(p')u_3(p) + u_2^*(p') u_4(p) + u_3^*(p') u_1(p) + u_4^*(p') u_2(p).
\end{aligned}
\end{equation}
Next, the positive and negative helicity eigenstates are given in Eq.(8.148), written as:
\begin{equation}\label{eq:spinors}
    u_{\uparrow} = N \begin{bmatrix} c \\ s e^{i\phi} \\ \alpha c \\ \alpha s e^{i\phi} \end{bmatrix}, \quad
    u_{\downarrow} = N \begin{bmatrix} -s \\ c e^{i\phi} \\ \alpha s \\ - \alpha c e^{i\phi} \end{bmatrix},
\end{equation}
where $\sin(\theta/2)=s$, $\cos(\theta/2)=c$, $\alpha = E/(p+m)$, and $N$ is a normalization. Without loss of generality, we set $\phi=0$, $p^{\mu}=(E,0,0,p)$, and $p'^{\mu}=(E,p\sin\theta,0,p\cos\theta)$. Thus, $c=1$ and $s=0$. The helicity-flip matrix amplitudes are
\begin{equation}\label{eq:M_up_down}
\begin{aligned}
M_{\uparrow \downarrow} &\propto \bar{u}_{\uparrow}(p') u_{\downarrow}(p)=|N|^2 \left[ c' \alpha s - s' \alpha c + \alpha c' (-s) + \alpha s' c \right] = 0,
\end{aligned}
\end{equation}
\vspace{-2.6em}
\begin{equation}\label{eq:M_down_up}
\begin{aligned}
M_{\downarrow \uparrow} &\propto \bar{u}_{\downarrow}(p') u_{\uparrow}(p) = |N|^2 \left[-s' \alpha c + c' \alpha s + \alpha s' c - \alpha c' s \right] = 0.
\end{aligned}
\end{equation}
The helicity-preserving amplitudes are
\begin{equation}\label{eq:M_same_spin}
M_{\uparrow \uparrow} = |N|^2 2 \alpha c', \quad
M_{\downarrow \downarrow} = |N|^2 (-2 \alpha c').
\end{equation}
The helicity flip amplitudes $\|M_{\downarrow \uparrow}\|^2$ and $\|M_{\uparrow \downarrow}\|^2$ are thus zero for a scalar exchanging interaction. For fermions with zero mass, their helicities are equivalent to chiralities, and the chirality flipping rate should be zero as well. However, for massive fermions, the chirality flipping rate is related to the overall scattering rate by the factor of $(m/3T)^2$. 

\nocite{*}
\bibliographystyle{apj}
\addcontentsline{toc}{chapter}{References}
\bibliography{mybib}

\end{document}